# Fungicides *vs* mycoinsecticides in the management of corn leafhopper: physicochemical, *in vitro* and *in vivo* compatibilities, and degradation kinetics in maize plants

Matheus Rakes[1*]; Maíra Chagas Morais[1,6]; Maria Eduarda Sperotto[2]; Odimar Zanuzo Zanardi[3]; Gabriel Rodrigues Palma[4]; Luana Floriano[5]; Renato Zanella[5]; Osmar Damian Prestes[5]; Daniel Bernardi[1]; Anderson Dionei Grützmacher[1]; Leandro do Prado Ribeiro[6]

[1]Department of Plant Protection, Faculty of Agronomy Eliseu Maciel, Federal University of Pelotas (UFPel), Pelotas, RS, Brazil

[2]Faculty of Agronomy, Uceff University (UCEFF), Chapecó, SC, Brazil

[3]Department of Education, Research and Extension, Federal Institute of Santa Catarina (IFSC), São Miguel do Oeste, SC, Brazil

[4]Hamilton Institute, Maynooth University, Maynooth, Kildare, Ireland

[5]Chemistry Department, Federal University of Santa Maria (UFSM), Santa Maria, RS, Brazil

[6]Research Center for Family Agriculture, Agricultural Research and Rural Extension Company of Santa Catarina (CEPAF/EPAGRI), Chapecó, SC, Brazil

*Corresponding author. E-mail address: matheusrakes@hotmail.com (M. Rakes)

**Running title:** Fungicides *vs* mycoinsecticides in the management of corn leafhopper

**Abstract**

The present study investigates the compatibility of mycoinsecticides based on isolates IBCB66 and Simbi BB15 of *Beauveria bassiana* and Esalq-1296 of *Cordyceps javanica*, which are registered for the management of *Dalbulus maidis* in Brazil, with synthetic fungicides. Irrespective of the fungicide, a total inhibition in the number of colony-forming units (CFUs), vegetative growth, conidiogenesis, and conidial viability of the three tested isolates was observed, with their incompatibility being indicated in the *in vitro* bioassays. However, the use of formulated mycoinsecticides mitigated the impact of these xenobiotics on the number of CFUs, with the commercial mycoinsecticide FlyControl® (*B. bassiana* isolate Simbi BB15) being the least sensitive to the fungicides propiconazole + difenoconazole, bixafem + prothioconazole + trifloxystrobin and trifloxystrobin + tebuconazole. Nevertheless, an increase in exposure time (from 1.5 to 3 hours) generally led to an increase in the toxicity of fungicides towards entomopathogens. Physical-chemical compatibility assessments indicated that physical incompatibilities were observed, depending on the mycoinsecticide formulation. In addition, *in vivo* bioassays employing *D. maidis* adults demonstrated that, despite a synergistic effect on mortality in certain binary mixtures, no cadavers exposed to such mixtures exhibited fungal extrusion. Furthermore, analyses using UHPLC/MS/MS revealed alterations in the degradation kinetics (*k*) of the active ingredient (a.i.) pyraclostrobin, with changes greater than tenfold being observed in the different formulations of the fungicides that were tested. Consequently, given the diminished degradation kinetics of the active ingredients in maize plants, the implementation of mycoinsecticides should precede, in isolation, the application of synthetic fungicides within the framework of phytosanitary management of maize crops.

**Keywords:** Entomopathogenic fungi, *Beauveria bassiana*, *Cordyceps javanica*, *Dalbulus maidis*, half-life

## Introduction

In recent years, the cyclical and structural changes in the agricultural scenario that have occurred due to climate change and the increase in demand for agri-food products, and the consequent intensification of land use, with up to three consecutive harvests in some tropical agricultural regions (*e.g.*, Brazil), have been identified as responsible for the increase in productivity losses caused by biotic factors, including phytopathogens and arthropod pests (Oliveira et al., 2016; Bahar et al., 2020; Mosier et al., 2021; Tonnang et al., 2022; Junaid and Gokce, 2024). In maize production systems, for example, these factors are considered accountable for the increased incidence and severity of leaf spots, stem and ear rots, and diseases transmitted by arthropod vectors, leading to production losses of up to 80% (Brito et al., 2011). Furthermore, given their high capacity for adaptation and favorable conditions for the development of their populations in tropical regions, several species of arthropod pests, mainly with sucking habits, have increased in abundance and caused significant damage to maize crops (Cruz et al., 2012; Panizzi et al., 2022; Cunha et al., 2023; Queiroz et al., 2025). In the last decade, the corn leafhopper, *Dalbulus maidis* (DeLong & Wolcott) (Hemiptera: Cicadellidae), has emerged as a major phytosanitary challenge in maize crops in the neotropical region, due to the potential direct and indirect damage caused (Oliveira & Frizzas, 2022), especially because it acts as a vector of pathogens (mollicutes and viruses) associated with the corn stunt complex (CSC) (Galvão et al., 2021; Albuquerque et al., 2022; Vilanova et al., 2022).

During the critical phase of *D. maidis* attack, which extends from emergence to the 10$^{th}$ fully expanded leaf ($V_{10}$) (Hanway, 1963), management has been predominantly reliant on the use of synthetic insecticides administered via seed treatments and foliar sprays (Pozebon et al., 2022). However, in recent harvests, microbiological insecticides based on entomopathogenic fungi (mycoinsecticides) have been widely adopted, representing approximately 15% of the global market (Rajula et al., 2021; Sharma et al., 2023). Among the mycoinsecticides utilized, those formulated with different isolates of *Beauveria bassiana* (Bals.-Criv.) Vuill. and *Cordyceps javanica* (Wize) (both Hypocreales: Cordycipitaceae) are used in approximately 2 million hectares of maize crops in Brazil (Parra, 2019; 2023). For the management of *D. maidis* alone, mycoinsecticides represent more than 95% of formulated products registered in the country (Agrofit, 2025). The efficacy in managing phytopathogens, notably those of a necrotrophic nature, is demonstrably associated with the timing of synthetic fungicide applications within the crop cycle. These applications are initiated in a preventative manner, particularly from the $V_4 – V_6$ phenological stages (4 – 6 fully expanded leaves), thereby ensuring the maintenance of the crop's productive capacity (Blandino et al., 2012; Wise et al., 2019; Kryuyi et al., 2020).

The application of insecticides and fungicides is often coincidental, and these products are commonly applied in combination in tank mixes. The primary advantage of this practice is the increased control spectrum and reduced application costs (Rakes et al., 2017; Raetano & Chechetto, 2019). According to Gazziero (2015), pesticides in tank mixtures are adopted in 95% of sprayings carried out by Brazilian farmers, using two to five products in the application mixture, with the most frequent association (31%) being fungicides with insecticides. Despite their extensive utilization, owing to the numerous advantages inherent in tank mixtures, particularly those comprising diverse types of pesticides, these mixtures are capable of inducing substantial alterations in the physicochemical characteristics of the mixture (Gandini et al., 2020; Rakes et al., 2018; 2024). Specifically, mixtures containing fungicides have been shown to result in significant reductions in the hydrogen potential (pH) of the mixture, primarily in dithiocarbamates and triazoles. These reductions can lead to chemical instabilities that alter the molecular structure, resulting in the formation of phytotoxic by-products that are detrimental to plant life (Carter and Wain, 1964; Kumar et al., 2014; Stefen et al., 2017). Additionally, these changes can reduce the efficacy of these products in the management of the target pest (Gandini et al., 2020). Furthermore, these characteristics have been shown to modify the degradation kinetics of pesticides, including synthetic fungicides (Tavares & Cunha, 2023), resulting in alterations to the foliar protection time when compared to isolated applications (An et al., 2024).

As demonstrated in several studies (Wu et al., 2020; Khun et al., 2020; Samal et al., 2024), the lethal effect of synthetic fungicides on entomopathogenic fungi has already been established. However, new active ingredients and formulations of both fungicides and mycoinsecticides have been developed and commercially released (Samal et al., 2024; Akoijam et al., 2024; Jain et al., 2024), mainly to increase environmental safety and the control spectrum. Nevertheless, to date, no study has evaluated the degradation kinetics of fungicides in their respective commercial formulations, which contain double or triple mixtures of active ingredients.

In this study, the primary hypothesis tested was that synthetic fungicides could be used concomitantly in tank mixtures with purified and formulated isolates of *B. bassiana* and *C. javanica*, which are registered for the management of *D. maidis* in Brazil, without causing deleterious effects on the secondary cycles of the epizootic. As a secondary hypothesis, we predict that the physicochemical compatibility of mixtures of fungicides and mycoinsecticides, as well as the half-life of the active ingredients of synthetic fungicides, influence their interaction with the mycoinsecticide, which may allow or restrict their joint application. To this end, we conducted *in vivo* and *in vitro* bioassays to evaluate the compatibility of nine widely used synthetic fungicides in maize crops in Brazil with the purified isolates IBCB66 and Simbi BB15 of *B. bassiana* and Esalq-1296 of *C. javanica*, or in two different commercial formulations. Furthermore, the physicochemical compatibility of binary mixtures between fungicides and mycoinsecticides was evaluated using the dynamic technique (with agitation). Finally, the degradation kinetics of fungicides after spraying on maize plant leaves were evaluated, and the half-life of different active ingredients in different commercial formulations was estimated.

## Materials and methods

### Fungicides and mycoinsecticides tested

Detailed information about the synthetic fungicides and mycoinsecticides used in the *in vitro* and *in vivo* bioassays are described in Table 1. The fungicides were evaluated at the maximum doses recommended by the manufacturers for maize crops in Brazil (Agrofit 2025). Furthermore, the mycoinsecticides were tested at the doses recommended by manufacturers and technicians in southern Brazil (personal information). The treatments were solubilized in distilled water, using a spray volume corresponding to 150 L $ha^{-1}$.

## *In vitro* bioassays

*In vitro* bioassays were conducted within a biological oxygen demand (BOD) chamber that had been calibrated to a temperature of 25 ± 1 ºC, a relative humidity (RH) of 75 ± 10%, and a photoperiod of 14 L: 10 D h. All bioassays were conducted under a completely randomized design.

## Obtaining and propagating entomopathogenic fungal isolates

The fungal isolates IBCB66 and Simbi BB15 from *B. bassiana* and Esalq-1296 from *C. javanica* were obtained through monosporic isolation from samples of commercial mycoinsecticides (Table 1). Following a purification process conducted within a vertical flow chamber, the isolates were propagated in Petri® dishes (Ø 9 cm) containing approximately 15 mL of culture medium based on potato dextrose agar [PDA Difco® (Becton-Dickinson Company, Franklin Lakes, New Jersey, USA)]. This medium was fortified with streptomycin at a concentration of 5 g $L^{-1}$.

## Vegetative growth of fungal isolates exposed to fungicide-treated culture media

To assess the detrimental impacts of synthetic fungicides on the vegetative growth of the respective entomopathogenic fungal isolates, the guidelines established by the International Organization for Biological and Integrated Control (IOBC) (Coremans-Pelseneer 1994) were implemented, with modifications outlined by Rakes et al. (2024).

The inoculum of each tested fungal isolates was obtained by scraping the surface of pure plates and propagating according to the previously described procedure. Thereafter, the inoculum was transferred to a solution containing 10 mL of sterile distilled water and 0.1% Tween 80® (v $v^{-1}$) and adjusted to a concentration of $1 \times 10^8$ conidia $mL^{-1}$. Subsequently, an aliquot of 200 µL was transferred to Petri dishes containing PDA culture medium (Difco®) and kept for 3 days in BOD chambers under the conditions mentioned above.

Prior to inoculation, the culture medium (Difco® PDA) was sterilized in Erlenmeyer® flasks for 15 min at a temperature of 121°C. Subsequently, the medium was cooled to a temperature of 45°C, when streptomycin, at a concentration of 5 g $L^{-1}$, and solutions containing the synthetic fungicides under investigation were incorporated (Table 1). Sterile distilled water was utilized as a negative control. Before solidification of the treated culture medium, a 15-mL aliquot of medium with the respective treatments was transferred to sterile Petri dishes (Ø 9 cm).

Following the solidification of the culture medium on the plates, colonies measuring 3 mm in diameter (obtained from the multiplied plates, as previously described) were inserted into the central point of each plate containing the medium treated with the various treatments (Table 1). For each treatment, 10 Petri dishes (replicates) were utilized. Following inoculation, the Petri dishes were placed in BOD chambers to assess the vegetative growth of the fungal isolates.

The assessment of vegetative growth occurred 12 days after the incubation of the treatments. The assessment entailed the measurement of two orthogonal dimensions of the colonies utilizing a 145 mm monobloc digital electronic caliper (Worker Tools Ltda., São José dos Pinhais, Paraná, Brazil). A 10 × 3 factorial scheme (treatments × fungal isolates) was employed for this bioassay.

**Conidiogenesis and viability of fungal isolates exposed to fungicide-treated culture media**

The number and viability of conidia produced were quantified by selecting five Petri dishes from each treatment at the endpoint (12 days) of the previous bioassay. To accomplish this, the plates were opened using a vertical flow chamber, and the colonies were meticulously excised with a scalpel and fine-tipped forceps, both of which were sterilized. Thereafter, the colonies were meticulously transferred to individual test tubes containing 10 mL of sterile distilled water and Tween 80® (0.1%, v/v$^{-1}$). Subsequently, each tube was shaken for 5 min in a vortex shaker (Kasvi Importação e Distribuição de Produtos para Laboratórios Ltda., São Paulo, SP, Brazil) at 2800 rpm to detach the conidia from the aqueous solution. The number of conidia produced (conidiogenesis) was subsequently quantified in a Neubauer chamber (Kasvi Importação e Distribuição de Produtos para Laboratórios Ltda., São Paulo, SP, Brazil) under an optical microscope (400 ×), based on serial dilutions.

To evaluate the viability of the conidia produced, a spore solution was adjusted to a concentration of $1\times10^6$ conidia mL$^{-1}$ for each plate selected in the previous assay. Thereafter, a 600-µL aliquot of each suspension was transferred to Petri dishes (Ø 9 cm) containing 10 mL of sterile bacteriological agar at a concentration of 15.6 g L$^{-1}$ (Kasvi Importação e Distribuição de Produtos para Laboratórios Ltda., São Paulo, SP, Brazil) and spread with the aid of a Drigalski loop. Conidial viability was quantified 14 and 18 hours after inoculation, according to the isolate tested, under an optical microscope (400 ×). For these bioassays (conidiogenesis and conidia germination), a 10 × 3 factorial scheme (treatments × fungal isolates) was employed.

## Compatibility between fungicides and pure isolates of entomopathogenic fungi

In order to synthesize the results obtained from the analysis of the parameters of vegetative growth, conidiogenesis, and conidial viability of the purified isolates after contact with the synthetic fungicides tested, the biological index (BI) proposed by Rossi-Zalaf et al. (2008) was used:

$$BI: [47 \times (VG) + 43 \times (SPO) + 10 \times GERM]/100$$

Where: *VG* is the percentage of vegetative growth compared to the control; *SPO* is the percentage of sporulation compared to the control; *GERM* is the percentage of conidial germination compared to the control. Thereafter, the results were classified as: incompatible ($0 \leq BI \leq 41$), moderately compatible ($41 < BI \leq 66$), and compatible ($BI > 66$).

## Effects of fungicides on colony-forming units (CFUs) of entomopathogenic fungi pure isolates

To verify the effect of the fungicides tested on the number of colony-forming units (CFUs) of purified isolates of *B. bassiana* (IBCB66 and Simbi BB15) and *C. javanica* (Esalq-1296), solutions of the respective isolates were initially standardized to a concentration of 500 spores mL$^{-1}$, in a volume of 10 mL of solution, containing the adhesive spreader Tween 80® a 0.1% (Sigma-Aldrich Corp., Sydney, Australia). The fungal suspension was quantified using a Neubauer chamber and a Zeiss AX10 microscope (Zeiss AG Corp., Oberkochen, Germany) equipped with a 400 × magnification lens. Subsequently, the fungal suspension was added to the solution containing the fungicides tested at the respective doses (Table 1) in Erlenmeyer® flasks with a capacity of 100 mL. The flasks were manually shaken, following the procedures described by Ribeiro et al. (2023) and Rakes et al. (2024). The flasks containing the treatments were then placed in BOD chambers at 25 ± 1ºC for a period of 1.5 and 3 hours of exposure.

After the exposure times, a 200-µL aliquot of each treatment was inserted into 15 mL of PDA Difco® culture medium plus streptomycin (5 g L$^{-1}$), arranged in sterile Petri dishes (Ø 9 cm), and spread with a Drigalski loop. An equivalent volume of sterile distilled water was utilized as a negative control. Following a 4-day incubation period, the number of colony-forming units (CFUs) was enumerated. Five Petri dishes (replicates) were utilized for each treatment level. This bioassay was conducted in a 10 × 3 × 2 factorial scheme (treatments × fungal isolates × exposure times).

**Effects of fungicides on the colony-forming units (CFUs) of mycoinsecticide commercial formulations**

A bioassay was carried out to quantify the number of colony-forming units (CFUs) to verify the sensitivity of the fungal isolates in their respective commercial formulations. To this end, two commercial mycoinsecticides were selected due to their different formulations: FlyControl® (OD formulation), formulated with the Simbi BB15 isolate of *B. bassiana*, and Octane® (CS formulation), containing the Esalq-1296 isolate of *C. javanica*. In Erlenmeyer® flasks, the treatments, which comprised binary mixtures of synthetic fungicides and mycoinsecticides, were added at the respective doses described in Table 1 in a volume of 100 mL. Subsequently, the treatments were manually agitated and maintained in BOD-type chambers at 25 ± 1ºC for exposure periods of 1.5 and 3 hours.

Following the exposure times (independent sampling), the treatment solutions (binary mixtures of mycoinsecticides and fungicides) were diluted 20-fold (with one part of the suspension containing the binary mixture of mycoinsecticide and fungicide to 19 parts of sterile distilled water). Thereafter, a 200-µL aliquot of the solution dilution for each treatment was inserted into sterile Petri dishes (Ø 9 cm), containing 15 mL of PDA culture medium (Difco®) plus 5 g $L^{-1}$ of streptomycin. An equivalent volume of sterile distilled water was utilized as a negative control. Following a 4-day incubation period, the number of colony-forming units (CFUs) was counted. Five Petri dishes (replicates) were used for each treatment level. A factorial scheme of 10 × 3 × 2 (treatments × fungal isolates × exposure times) was employed.

**Physicochemical compatibility**

The physicochemical compatibility of the binary mixtures of fungicides and mycoinsecticides was conducted in a climate-controlled room (temperature: 25 ± 1ºC and 65 ± 10% relative humidity (RH)). The chemical characteristics and the stability of the mixture were evaluated according to the recommendations of the Brazilian Standard (NBR) 13875:2014 (ABNT 2014), with adaptations to the order of addition of the products described by Rakes et al. (2018), Ribeiro et al. (2023), and Morais et al. (2024).

The mixtures containing synthetic fungicides, mycoinsecticides, or binary mixtures of both were prepared using graduated glass beakers and water with a hardness of 20 mg $kg^{-1}$ $CaCO_3$. Initially, 150 mL of standard water and the respective fungicides/mycoinsecticides were added to each beaker. Subsequently, the test tube was sealed with plastic film and rotated 180° ten times. Finally, an additional 100 mL of standard water was added, resulting in a total volume of 250 mL of mixture per treatment. Subsequently, the mixtures were transferred to Becker® type cups with a capacity of 400 mL and kept under stirring for 2 hours on an IKA® RT-15 magnetic stirrer (IKA Brasil Equipamentos Laboratoriais, Analíticos e Processos Ltda., Campinas, SP, Brazil). Following the stirring process, the mixtures were allowed to rest for 10 min to facilitate the evaluation of their chemical and physical characteristics.

Chemical analyses were conducted using the hydrogen potential (pH), employing a bench pH meter (Digimed Instrumentação Analítico Ltda., São Paulo, SP, Brazil), and evaluating the electrical conductivity (EC) of the mixtures, utilizing an Orion 5-Star conductivity meter (Thermo Fischer Scientific Corp., Waltham, Massachusetts, USA). Physical analyses were conducted through visual examination, assessing the presence (P) or absence (A) of homogeneity, flocculation, sedimentation, phase separation, oil separation, crystal formation, cream formation, foam formation, lump formation, and the presence of lumps larger than 149 µm (ABNT, 2010).

***In vivo* bioassays**

Adults of *D. maidis* were obtained from a laboratory colony established with specimens collected in cornfields located in the municipality of Chapecó, Santo Catarina State, Brazil (27°05'16.0" S; 52°38'14.4" W). The entire process of rearing and maintenance of leafhoppers was carried out in air-conditioned rooms at a temperature of 25 ± 1°C, 60 ± 10% relative humidity (RH), and a photoperiod of 14 L: 10 D hours. The maintenance of laboratory-reared leafhoppers was accomplished through the use of plastic vessels, each containing approximately 0.4 L of sterilized substrate. This substrate was composed of a blend of *Pinus* sp. bark, peat, and expanded vermiculite (Tropstrato®; Vida Verde Indústria e Comércio de Insumos Orgânicos Ltda., Mogi Mirim, SP, Brazil). Maize plants (hybrid P3016VYHR) (Corteva Agriscience do Brasil Ltda., Barueri, SP, Brazil) in $V_3$ – $V_4$ phenological stages (Hanway 1963) were utilized as a feeding and oviposition substrate. The rearing of insects was placed in cages (35 × 30 × 30 cm in height, length, and width, respectively) made from plastic frames and covered with anti-aphid mesh (200-mm mesh). The maize plants were changed every two days until the insect's life cycle was completed.

For *in vivo* bioassays, adult, unsexed insects, 5-7 days old, were utilized. Prior to the application of the treatments (synthetic fungicides and isolated mycoinsecticides, as well as their respective binary mixtures), the insects were anesthetized with $CO_2$ for a duration of 5 sec. They were then placed in a dorsal position in the center of Petri dishes. The treatments were applied via an airbrush (OVD Importadora e Distribuidora Ltda., Novo Hamburgo, RS, Brazil) equipped with a 0.3 mm spray tip, a working pressure of 20 PSI, and an application volume of 1 mL, calibrated to apply the equivalent of 150 L $ha^{-1}$ of spray volume. Subsequent to the application of each treatment, a total of 10 insects were transferred to plastic cages (700 mL) containing a maize plant at the $V_2$ phenological stage, with the upper part sealed with thin *voile*-type fabric. On the substrate, a sterile aluminum paper disc was placed to avoid contamination from dead insects. The mortality assessments were conducted on a daily basis for a period of 9 days, as previously described by Rakes et al. (2024). Insects that did not show any movement when stimulated with a fine-tipped brush were considered dead. Five experimental units (replicates) were utilized for each treatment, with 10 insects per replicate ($n$ = 50).

On a daily basis, dead insects were extracted from the cages and sterilized with sodium hypochlorite (1%, v $v^{-1}$) for 20 sec. Thereafter, they were immersed in sterile distilled water three times. Subsequently, each insect was transferred to a sterile cryogenic tube (2 mL), containing hydrophilic cotton soaked with sterile distilled water, to maintain humidity inside the tube. The cryogenic tubes were then placed in an incubator for a period of 10 days, with the objective of observing fungal extrusion in the cadavers. Prior to conidia quantification on cadavers, extruded fungi were identified using the morphological characters described by Barnett and Hunter (1972). Thereafter, 1 mL of sterile distilled water and Tween 80® at a concentration of 0.1% (v $v^{-1}$) were added to the cryogenic tubes and stirred in a vortex shaker for 5 min at 2800 rpm. The suspensions were then serially diluted, and the conidia from the samples were quantified in a Neubauer chamber under an optical microscope (400 ×). All *in vivo* bioassays were conducted in BOD chambers maintained at 25 ± 1°C, 80 ± 10% relative humidity (RH), and a photoperiod of 14 L:10 D hours.

## Analysis of fungicide degradation kinetics in maize leaves

Maize plants (hybrid P3016VYHR) were cultivated in plastic pots (25 L capacity) containing soil classified as Dystrophic Red Latosol (Potter et al., 2004), with the following characteristics: clay = 32 % (w $v^{-1}$); water pH (1:1) = 5.4; P = 28.0 mg $dm^{-3}$; K = 336.0 mg $dm^{-3}$; organic matter = 4.9% (m $v^{-1}$). The plants were maintained within a greenhouse, equipped with an artificial ventilation system and a 10-mm alveolar polycarbonate cover. The polycarbonate's average transmittance to global solar radiation was 70%. The experiment recorded the following temperature values: 31.3°C, 13.9°C, and 25.8°C for maximum, minimum, and average, respectively. The mean relative humidity was 59%, with a maximum of 94.7% and a minimum of 40.4%. The temperature and relative humidity data were recorded at 30-minute intervals throughout the test run using a Rotronic HL-1D data logger (Rotronic AG, Bassersdorf, Switzerland). Luminosity was measured using an LDV 2000 digital lux meter (OVD Importadora e Distribuidora Ltda., Novo Hamburgo, RS, Brazil). The maximum recorded luminosity was 160.1 × 100 lux, the minimum was 53.0 × 100 lux, and the average was 90.8 × 100 lux.

The synthetic fungicides were applied at the $V_6$ phenological stage of plant development, employing a Guarany® Ultrajet 500 mL manual sprayer until the point of runoff. The plants were maintained outdoors for a duration of 15 min to allow the residual mixture to evaporate. The plants were sprayed with distilled water in the negative control. The experiment was conducted using 9 pots per treatment, with each pot containing a single plant. The pots were arranged in a completely randomized design.

Leaf sampling was conducted on a periodic basis at 0 (6 h after treatment application), 2, 4, 8, 16, and 24 days after treatment application (DAAT). Each sample consisted of the extraction of approximately 20 g of leaves (which had received the application) on a random basis in the repetitions of each treatment. The samples were then subjected to immediate identification, followed by transportation to the designated laboratory. Thereafter, the samples were stored in an ultra-low temperature freezer (Panasonic VIP Plus+, Panasonic do Brasil Ltda., São Paulo, SP, Brazil) at -80°C for subsequent analysis.

The following reagents were utilized for chromatographic analyses (analysis grade): acetonitrile, formic acid, ammonium formate, anhydrous magnesium sulfate, and sodium chloride, obtained from J.T. Baker (Phillipsburg, NJ, USA). The solid standards of the active ingredients tested (Table 1) were obtained from LGC Standards (Augsburg, Germany), with the highest purity available. Individual stock standard solutions at 1000 mg $L^{-1}$ and an intermediate working mix at 10 mg $L^{-1}$ were prepared in acetonitrile and stored at -18ºC.

The preparation of the samples followed the procedures outlined by Viera et al. (2017). The homogenization of the plant material was conducted using a VariMix model food processor (Varimix Comercio & Servicos Ltda., Itajai, SC, Brazil) with dry ice, with the objective of preventing the degradation of the compounds. The sample preparation procedure was based on the QuEChERS (Quick, Easy, Cheap, Effective, Rugged, and Safe) method. Initially, 3 g of leaf sample were weighed in a 50 mL Falcon-type polypropylene (PP) tube (CRAL Artigos para Laboratório Ltda., Cotia, SP, Brazil). Subsequently, 10 mL of acetonitrile containing 5% (v $v^{-1}$) formic acid was added, and the tube was shaken in a vortex for 1 min. For the partitioning step, a mixture of 1.5 g of NaCl and 4 g of $MgSO_4$ was used, and the tube was vortexed for 1 min and subsequently subjected to centrifugation at 2600 g for 8 min. In the cleaning step, 2 mL of the upper layer was transferred to a 15 mL polypropylene tube containing 300 mg of $MgSO_4$, 50 mg of C18, and 10 mg of graphitized carbon black (GCB). Subsequently, the tube was shaken in a vortex for 1 min, followed by centrifugation at 2600g for 8 min. Finally, the extract was filtered through a 0.2 µm nylon syringe filter and diluted 5-fold with ultrapure water prior to analysis by ultrahigh performance liquid chromatography coupled to tandem mass spectrometry (UHPLC-MS/MS).

The active ingredients of each synthetic fungicide were quantified by the UHPLC-MS/MS technique on a Waters system (Waters Corporation, Milford, MA, USA) equipped with an Acquity UPLC™ binary pump liquid chromatograph, Xevo TQ™ MS/MS triple quadrupole mass spectrometer, autosampler, column temperature controller, and MassLynx V4.1 data acquisition software. The nitrogen gas was obtained from a Peak Scientific generator (Peak Scientific Instruments Ltd, Inchinnan, Scotland, UK) model NM30L-MS, and argon gas (6.0) was used as the collision gas. For chromatographic separation, an Acquity UPLC™ BEH C18 column (50 × 2.1 mm, 1.7 µm particle size) was utilized, with temperature maintained at 40°C. The mass spectrometer was operated in selected reaction monitoring (SRM) mode, utilizing the most intense transition for quantification and the second most intense transition for identification. The optimization of mass spectrometry (MS) parameters was achieved through the direct infusion of individual pesticide solutions into the mass spectrometer.

The UHPLC-MS/MS conditions were as follows: capillary voltage 2 kV, desolvation temperature 500°C, nitrogen flow rate (desolvation gas) 600 L $h^{-1}$, spray flow rate 80 L $h^{-1}$, argon flow rate (collision gas) 0.15 mL $min^{-1}$, source temperature 150 °C, and injection volume 10 µL, as proposed by Rizzetti et al. (2016) and Viera et al. (2017). The instrument was operated using an electrospray ionization (ESI) source in positive mode. The mobile phase comprised water (A) and methanol (B), both containing 0.1% (v $v^{-1}$) formic acid and 5 mmol $L^{-1}$ ammonium formate. The mobile phase gradient commenced at 5% B and remained constant until 0.25 min, when it increased linearly to reach 100% B in 7.5 min (maintained for 0.75 min). Thereafter, it returned to 5% B in 8.81 min (maintained for 1.19 min). The flow rate was kept at a constant rate of 0.25 mL $min^{-1}$, and the total duration of the experiment was 10 min.

For the purpose of quantification, analytical curves were prepared from standard solutions for each active ingredient, which had been prepared in the white extract of the matrix. To evaluate the procedure developed for analyzing pesticide residues in maize leaves, the method was validated following the guidelines of Sante (2015) for the parameters selectivity, matrix effect, linearity, limits of detection (LOD), limit of quantification (LOQ), and precision (intraday and interday). Selectivity was evaluated by comparing the chromatograms obtained by injections of maize leaf extracts devoid of pesticide residues. The accuracy and precision (intraday and interday) of the method were evaluated in terms of recovery assays (70–120% recovery) and, for precision, with the relative standard deviation (RSD ≤ 20%). The LOQ of the method was established as the lowest fortification levels that demonstrated adequate accuracy and precision. The LOD of the method was obtained by dividing the LOQ value by 3.

**Data analysis**

For the *in vitro* bioassays, a Poisson generalized linear model (GLM) (McCullagh and Nelder 1989; Demétrio et al. 2014) was fitted to the colony-forming units (CFUs), incorporating the effects of treatment, time, and the two-way interaction between treatment and time in the linear predictor as fixed. Conversely, for the *in vivo* bioassays, a binomial GLM was fitted to the mortality data, including the treatment effects in the linear predictor as fixed.

The time-until-event variable (Death) was analyzed by fitting Cox proportional-hazards models, including the effects of the treatments and a gamma frailty to account for correlations between observations taken within the same experimental unit in the linear predictor (log link). A negative binomial GLM was fitted to the number of conidia on the cadavers data, including the effects of treatment on the linear predictor as fixed.

The foliar degradation kinetics of each active ingredient (a.i.) present in the studied fungicides was determined from the active ingredient plot as a function of time, following the procedure outlined by Boesten et al. (2006). A linear model was fitted to the log of the total amount of active ingredients including the effect of time to estimate the parameter k of a simple first-order (exponential decay) model (SFO):

$$\mathrm{M} = \mathrm{M}_0 \mathrm{e}^{-\mathrm{kt}}$$

Where: M is the total amount of active ingredient at time t; M0 is the total amount of active ingredient present at time t=0 and k is the relative degradation rate. The half-lives of each compound concentration were calculated with the following equation:

$$\mathrm{DT}_{50} = \frac{\ln(2)}{k}$$

The significance of effects, including isolated effects and multiple interactions was assessed through likelihood ratio (LR) tests for nested models. For GLMs, the goodness-of-fit was assessed using half-normal plots with a simulated envelope (Moral et al. 2017) and for Cox proportional hazards models, the goodness-of-fit was assessed by visual analysis of the martingale residuals (Therneau and Grambsch 2000). All statistical analyses were performed using the R environment software (R Core Team 2021).

## Results

### *In vitro* compatibility

### Vegetative growth, conidial production and viability of fungal isolates exposed to fungicide-treated culture media

Following a 12-day incubation period, a complete inhibition was observed when evaluating the diameter of colonies of different entomopathogenic fungal isolates developed in PDA culture medium treated with the tested fungicides (Table 2). Consequently, the assessment of conidiogenesis and conidial viability of fungal isolates exposed to fungicides-treated culture media was not feasible (Table 2). In general, all tested isolates exhibited a similar vegetative growth in the negative control (distilled water), while the number of conidia produced was increased in the isolates Esalq-1296 of *C. javanica* and IBCB66 in the absence of fungicides in the PDA media (negative control; Table 2). However, conidial viability was greater than 97% in all fungal isolates in the negative control (Table 2).

Furthermore, the biological index (*BI*) proposed by Rossi-Zalaf et al. (2008) indicated that all fungicides tested were considered incompatible with the pure isolates of Esalq-1296 of *C. javanica* and IBCB66 and Simbi BB15 of *B. bassiana* (Table 3).

### Effects of fungicides on colony-forming units (CFUs) of entomopathogenic fungi pure isolates and commercial formulations

Irrespective of the exposure time (1.5 and 3 h), all fungicides tested completely inhibited the number of colony-forming units (CFUs) of pure isolates of three fungal isolates tested in a bioassay method simulating a tank mixture (Table 4). In the negative control, the highest CFU number was observed in isolate IBCB66 from *B. bassiana*, followed by isolate Esalq-1296 from *C. javanica* (Table 4).

Conversely, the formulation procedure used in two tested commercial formulations [oil dispersion (OD; FlyControl®) and concentrate suspension (CS; Octane®)] mitigated the negative effect of some fungicides (Table 5). After a 4-day incubation period, a significant interaction between fungicides × exposure times was observed when evaluating the number of CFUs of FlyControl® ($\chi^2$= 533.41; df = 8; $p$ < 0.01) and Octane® ($\chi^2$= 252.18; df = 4; $p$ < 0.01) (Table 5).

For FlyControl® (*B. bassiana* isolate Simbi BB15), all fungicides (except propiconazole +difenoconazole and difenoconazole + pydiflumetofen at 1.5 h) exhibited a reduction in the number of CFUs compared to the negative control, regardless of the exposure time (Table 5). Except for the treatment with the pyraclostrobin + epoxiconazole fungicide (total inhibition), increasing the exposure time (from 1.5 to 3 h) promoted a significant reduction in the number of CFUs of this mycoinsecticide (Table 5). The same trend was observed for Octane® (*C. javanica* isolate Esalq-1296), with all treatments promoting a significant reduction in the number of CFUs compared to the negative control (Table 5). In addition, increasing the exposure time promoted a significant reduction of the treatments with an expressive number of CFUs (Table 5). In general, mycoinsecticide Octane® (CS formulation) showed a high sensitivity to the tested fungicides compared to FlyControl® (OD formulation).

**Physicochemical analysis**

Synthetic fungicides and mycoinsecticides alone exhibited pH values ranging from 4.74 to 7.07 and a electrical conductivity (EC) values ranging from 6.46 to 38.70 (Table 6). In general, binary mixtures containing fungicides and mycoinsecticides did not demonstrate significant changes in the pH and EC values (Table 6). In addition, binary mixtures containing fungicides based on epoxiconazole + fluxapyroxad + pyraclostrobin, bixafem + prothioconazole + trifloxystrobin, trifloxystrobin + tebuconazole and propiconazole + difenoconazole with the mycoinsecticide FlyControl®, as well as epoxiconazole + fluxapyroxad + pyraclostrobin + Octane®, showed no homogeneity and the presence of phase separation at the end of the stirring period (10 min) (Table 6).

***In vivo* bioassays**

After 9 days of topical exposure of *D. maidis* adults to synthetic fungicides and mycoinsecticides applied alone or in binary mixtures, significant differences were observed in the percentage of insect mortality ($\chi^2$ = 262.21; df = 29; $p$ < 0.01) (Table 7). In addition to the positive control (total mortality), the highest mortality rate (64%) was observed in the treatment with the fungicide based on epoxiconazole + fluxapyroxad + pyraclostrobin in combination with the fungal isolate Simbi BB15 of *B. bassiana* (FlyControl®), which differed significantly from the negative control treatment (Table 7). In addition, all treatments (synthetic fungicides and mycoinsecticides) applied in isolation caused mortality of less than 28% and were therefore not significantly different from the negative control (Table 7).

Significant differences were observed in the survival probability of *D. maidis* adults exposed to synthetic fungicides and mycoinsecticides applied alone (Figure 1A; $\chi^2 = 295.31$; df = 10; $p < 0.01$) or in binary mixtures with synthetic fungicides and mycoinsecticides FlyControl® (Figure 1B; $\chi^2 = 88.84$; df = 9; $p < 0.01$) and Octane® (Figure 1C; $\chi^2 = 165.94$; df = 9; $p < 0.01$). In general, only the fungicide combination of epoxiconazole + fluxapyroxad + pyraclostrobin reduced the probability of survival of *D. maidis* by more than 25% (Figure 1A). When fungicides were mixed with any of the mycoinsecticides tested, the greatest reductions in survival probability occurred immediately to one day after application (Figure 1B, C). In addition, when commercial mycoinsecticide were applied alone, there were significant reductions in survival probability after 4 days of leafhopper contamination (Figure 1B, C).

Of the dead adults of *D. maidis* collected and maintained in conditions conducive to the development of entomopathogens, only those exposed to treatments with the mycoinsecticides FlyControl® and Octane® applied alone exhibited fungal extrusions in 21.43 and 27.27% of the insect cadavers, respectively (Table 7). In addition, a significant difference was observed in the number of conidia produced per dead leafhopper ($\chi^2 = 16.89$; df = 1; $p < 0.01$), with the mycoinsecticide Octane® producing a higher number of conidia per cadaver (Table 7).

**Degradation kinetics of fungicides in maize plants**

Validation of the analytical method showed that the LOQ was 0.017 mg kg$^{-1}$ and the LOD was 0.005 mg kg$^{-1}$ for the active ingredient epoxiconazole. For all other compounds, the LOQ was 0.008 mg kg$^{-1}$ and the LOD was 0.005 mg kg$^{-1}$ (data not shown).

The concentration of residues of active ingredients applied to maize plants in the respective commercial formulations used decreased significantly with time after application (Figure 2 A-I). At 0 DAA, the quantification of the residues of the active ingredients varied from 37,250.23 mg kg$^{-1}$ for tebuconazole contained in the fungicide Nativo® (Figure 1E) to 1,154.45 mg kg$^{-1}$ for prothioconazole contained in the fungicide Fox Xpro® (Figure 1C). In contrast, residues in maize leaves at 24 DAA ranged from 12,405.41 mg kg$^{-1}$ for tebuconazole from the fungicide Nativo® (Figure 1E) to 103.75 mg kg$^{-1}$ for pyraclostrobin from the fungicide Abacus HC® (Figure 1A).

Adjusting the first-order degradation model (SFO), individually for the active ingredients within each commercial product studied, substantial differences were observed regarding the relative degradation kinetics (parameter *k*) (Figure 1 A-I; SM – Table 1). For instance, the active ingredient pyraclostrobin showed differences in degradation kinetics within the commercial fungicides Abacus HC® (Figure 1A), Ativum® (Figure 1B) and Orkestra® (Figure 1F), with *k* values of 0.206, 0.032, and 0.016, respectively (SM – Table 1). Based on this, the results regarding half-life time ($DT_{50}$) varied from 52.97 days (difenoconazole – Figure 1D) to 3.36 days (pyraclostrobin – Figure 1A).

**Discussion**

The *in vitro* bioassays indicated the high toxicity of the tested fungicides on the number of colony-forming units (CFU), vegetative growth, conidiogenesis, and viability of entomopathogenic fungal isolates when tested purified (not formulated). Moreover, while the fungal isolates in their respective formulations (FlyControl® - Oil-based (OD) formulation or Octane® - concentrated suspension (CS) formulation) exhibited, in general, reduced sensitivity upon contact with fungicides, and the *in vivo* bioassays revealed an increase in the mortality of *D. maidis* in some binary mixtures. No extrusion of conidia was observed in the insect cadavers exposed to binary mixtures between fungicides and mycoinsecticides. This condition could potentially compromise the secondary cycles of these entomopathogens in the field. Additionally, our findings suggest that the half-life times ($DT_{50}$) of the active ingredients (a.i.) vary across different commercial formulations of the fungicide in which these active ingredients are employed.

From a broad perspective, the cycle of an entomopathogenic fungus initiates with the germination of virulent conidia within the host's cuticle, a phase that is imperative for the infection success (Islam et al., 2021). Previous studies have identified the deleterious effects of fungicides on the germination of entomopathogenic conidia, caused by a fungistatic effect or even by the death of the fungus' reproductive structure (Qayyum et al., 2021). Celar and Kos (2020) concluded that copper-based protective fungicides reduce the germination of the *B. bassiana* isolate ATCC74040 by more than 90%. Furthermore, fungicides based on difenoconazole, propiconazole, trifloxystrobin, and azoxystrobin caused a fungistatic effect on *Metarhizium anisopliae* (Metch.) Sorok (Ascomycota: Clavicipitaceae), completely inhibiting the conidial germination process for more than 20 hours after exposure (Silva et al., 2013). In corroboration with these studies, our results indicated a total inhibition of the germination process of purified (non-formulated) conidia of *B. bassiana* and *C. javanica* when exposed to sprays containing the fungicides tested for 1.5 and 3 hours, simulating a tank mixture typically used in the field. This inhibition was measured through the number of colony-forming units (CFUs).

Despite the predominance of mycoinsecticide formulations in the Brazilian market, which are primarily of the wettable powder (WP; dry conidia) type, these formulations have been utilized for purposes such as enhancing spray stability, extending shelf-life, and protecting against extreme environmental factors, and mitigating the deleterious effects of pesticides in tank mixes (Lopes et al., 2011; Oliveira et al., 2018; Zaki et al., 2020; Behle and Birthisel, 2023). The findings of this study demonstrated that, depending on the exposure time and the synthetic fungicide under investigation, OD and CS formulations influenced the reduction in sensitivity of the respective fungal isolates in mixture with some of the synthetic fungicides tested. Moreover, in general, mycoinsecticide FlyControl®, formulated with an oil base (OD formulaiton), exhibited a reduced sensitivity to certain fungicides when compared to mycoinsecticide Octane® (CS formulation). These findings are consistent with those reported by Khun et al. (2020), who determined that conidia suspended in oil emulsions are partially protected from the harmful action of the fungicide pyraclostrobin. Additionally, Rakes et al. (2024) found that the mycoinsecticide based on the *B. bassiana* isolate Simbi BB15 (OD formulation) exhibited reduced sensitivity when exposed to various post-emergent herbicides utilized in maize crops in Brazil. However, despite such protection, our results indicated that, regardless of the mycoinsecticide formulation, the increase in exposure time (from 1.5 to 3 h) significantly reduced the number of colony-forming units (CFUs). This indicates that, in practice, the time between mixture preparation and actual spraying should be as short as possible when such mixtures are used, since significant reductions in compatibility indices can be observed (Ribeiro et al., 2023; Rakes et al., 2024; Silva et al., 2024).

In addition to interactions related to biological factors, it has been documented that the utilization of a single product in the spray tank is employed by only 3% of Brazilian producers (Gazziero, 2015). Consequently, despite the paramount importance of utilizing quality water as a carrier medium (Daramola et al., 2022), incompatibilities may arise, potentially undermining the efficacy of spraying operations (Gandini et al., 2020). In this context, our study revealed no significant alterations in the pH and electrical conductivity (EC) values of binary mixtures between fungicides and mycoinsecticides, when compared to the parameters of the products tested individually. Consequently, these mixtures can be classified as chemically compatible. In the context of fungicides, particularly systemic ones, it is imperative to employ optimal pH and EC ranges, as these parameters are directly associated with the absorption and translocation of the active ingredient within the plant (Wang et al., 2021; Li et al., 2022). Furthermore, the composition of mixtures can lead to alterations in rheological characteristics, including viscosity, surface tension, and droplet size of sprays. This, in turn, can result in uneven depositions of drops on the leaf canopy (Basi et al., 2012; Carvalho et al., 2017; Cunha et al., 2017; Zandonandi et al., 2018).

Gazziero (2015) also reported that in more than 70% of cases of physical incompatibilities of pesticide mixtures, the spray tips become clogged. The findings of this study indicated a lack of homogeneity in certain treatments comprising binary mixtures of fungicides with the mycosesticides FlyControl® and Octane®. However, the observed heterogeneity was attributed to phase separation, which was evident after an agitation period of 10 min. This finding suggests that, in practical terms, this issue can be readily addressed by maintaining constant agitation during the preparation and application of the solution.

Following the successful penetration of the target insect cuticle by the entomopathogen, the host colonization process is initiated through mycelial growth (Islam et al., 2021; Samal et al., 2024; Silva et al., 2024). This phase occurs inside the host body and is therefore less exposed to pesticides. However, it is important to note that this stage of the biological cycle precedes conidiogenesis, which is responsible for entomopathogen multiplication, ensuring the success of the epizootic in the field (Ribeiro et al., 2014; Sharma and Sharma, 2021). Consequently, our findings indicate that the insertion of a colony (Ø 3 mm) in the center of plates containing culture medium treated with the tested fungicides resulted in the complete inhibition of vegetative growth and, consequently, conidiogenesis of the purified isolates of *B. bassiana* and *C. javanica*. In their study, Khun et al. (2020) determined that the active ingredient pyraclostrobin (not formulated) is detrimental to purified conidia of *B. bassiana* isolate PPRI5339, even at doses 16-fold lower than those recommended for field use. In a similar line, D'Alessandro et al. (2011) found that azoxystrobin significantly reduced radial growth and conidiogenesis of *C. javanica* isolate CEP304 (formerly *I. fumosorosea*). Conversely, Shah et al. (2009) documented that azoxystrobin exerted no effect on the mycelial growth of *I. fumosorosea* isolate PFR97. To date, most studies conducted have quantified the toxicity of fungicides with only one active ingredient in their formulation on different isolates of entomopathogens (Samal et al., 2024). However, these data do not reflect reality, as in Brazil alone, more than 60% of fungicides registered for maize crops contain two or more active ingredients in the same formulation (Agrofit, 2025).

The utilization of disparate products in tank mixtures has the potential to elicit novel interactions with target arthropods that may not otherwise be discerned in *in vitro* bioassays (Gandini et al., 2020). Consequently, *in vivo* bioassays were conducted to provide a more comprehensive evaluation of these interactions. The findings of this study suggested a potential synergistic effect of certain fungicide and mycosinsekticide combinations in assessing the mortality of *D. maidis* adults. Furthermore, the highest mortality rates were observed in mixtures containing the oil-based mycosinsecticide FlyControl® (OD formulation). In general, both inerts in oil-based formulations and adjuvants increase the electron-dense regions of the cuticle, evidencing ruptures (Arnosti et al., 2019). Additionally, many oils have been observed to obstruct spiracles, thereby impeding gas exchange. In mixtures with other pesticides, including both biological and chemical agents, these oils have been found to enhance the adhesion of active ingredients to the cuticular layers (Ortiz-Urquiza and Keyhani 2013; Jarrold et al. 2007; Islam et al. 2021).

However, the findings of the present study demonstrated that, while the highest mortality rates of *D. maidis* occurred in specific binary mixtures that were examined, extrusion was not observed in any of the dead insects. This finding has the potential to compromise the efficacy of microbial control with entomopathogenic fungi under field conditions, as it hinders the occurrence of secondary cycles of the epizootic, thereby diminishing the “residual effect” of the strategy and its contribution to the balance of agroecosystems (Qayyum et al., 2021). Furthermore, our results demonstrated that, even when mycosesticides registered for the management of *D. maidis* in Brazil were applied to corn leafhoppers (topical contact), mortality rates below 30% were observed, and less than 30% of the cadavers showed fungal extrusion. In a recent study, Lopes et al. (2024) observed that *D. maidis* adults engage in a self-cleaning behavior, which has the effect of reducing the amount of inoculum and the lethality of mycoses. The authors also reported the excretion of a protein-lipid fluid produced in the Malpighian tubules, called brochosomes, which *D. maidis* uses to coat its body and protect itself from environmental stressors, including xenobiotics.

The integration of synthetic pesticides and mycosesticides into a pest management program necessitates a profound comprehension of the potential repercussions of pesticides on biological agents (Khun et al., 2020) in order to ascertain the optimal application timing (preceding, in tank mixtures or following) within a stipulated crop management context. In light of the aforementioned considerations, the findings of the present study, derived from both *in vitro* and *in vivo* bioassays, indicate that the synthetic fungicides and mycosesticides under evaluation should not be applied concurrently in tank mixtures, despite the operational advantages that might be offered. As Griffin (1994) and Tomlin (1997) have previously observed, the majority of fungicides possess a comparatively brief foliar persistence, typically exhibiting a half-life ($DT_{50}$) of less than 7 days. However, it is acknowledged that the degradation kinetics of pesticides are closely related to the type of pesticide, climatic conditions, plant metabolism (plant matrix), leaf age, and the mixtures used in the spray tank (Jacobsen et al., 2015; Viera et al., 2017; Rakes et al., 2021; Tavares and Cunha, 2023). However, to date, no study has quantified the degradation kinetics of fungicides used in maize crops. Consequently, this study employed UHPLC-MS/MS analyses to assess the foliar degradation kinetics of each active ingredient (a.i.) present in the formulations of the commercial fungicides under evaluation. The findings revealed that the median degradation time ($DT_{50}$) of a given active ingredient exhibited significant variation across different formulations. Furthermore, a relationship with the active ingredient concentration was not observed, since, in some cases, lower active ingredient concentrations presented higher $DT_{50}$. These findings are consistent with the observations reported by Rakes et al. (2021), who noted no significant changes in the $DT_{50}$ of azoxystrobin (approximately 17 days) when studied at concentrations of 250 and 500 mg a.i. $L^{-1}$. Consequently, based on the analyses conducted, it is recommended that the application of mycosinecticides intended for the management of *D. maidis* precedes the application of fungicides in maize crops. These findings are consistent with those reported by Khun et al. (2020), who concluded that pyraclostrobin residues should be reduced by more than 93% prior to the application of entomopathogens. Despite the elevated $DT_{50}$ values determined in this study, the application of entomopathogens alone is expected to result in the death of *D. maidis*, thereby leading to the extrusion of cadavers and subsequent dispersion of spores in the atmosphere. In the event that this hypothesis is validated under field conditions, *D. maidis* nymphs, which primarily inhabit the abaxial surface of leaves (an area where fungicide residues are unlikely to be present), will become infected. Through vertical transmission, secondary cycles of epizootics may be established, thereby contributing to the balance of agroecosystems and enhancing the efficacy of *D. maidis* control in cornfields.

**Authors' contributions**

MR: Conceptualization, Methodology, Investigation, Data Curation, Formal Analysis, Writing – Original Draft. MCM, MES, LF, RZ and ODP: Investigation, Data Curation, Writing – Review & Editing. GRP: Data Curation, Formal Analysis, Writing – Review & Editing. OZZ: Conceptualization, Writing – Original Draft and Writing – Review & Editing. DB, ADG and LPR: Conceptualization, Resources, Supervision,

Project administration, Funding acquisition, Writing – Review & Editing. All authors read and approved the final version of this manuscript.

## Declarations

### Funding

The authors wish to thank the Brazilian National Counsel of Technological and Scientific Development (CNPq, grant 310385/2022-9), Foundation for Research and Innovation of the State of Santa Catarina (FAPESC, grant 2023TR000220), and Alfa Agroindustrial Cooperative (CooperAlfa) for the financial support. We also thank the CNPq (process 163544/2021-2) for scholarship conceives to the first author. Finally, we thank Taighde Éireann – Research Ireland under Grant 18/CRT/6049 for the financial support. The opinions, findings and conclusions or recommendations expressed in this material are those of the authors and do not necessarily reflect the views of the funding agencies.

### Conflict of interest

The authors declare no conflict of interest.

### Availability of data and material

Not applicable

### Code availability

Not applicable

### Ethical approval

This study does not contain any studies with human participants or large animals performed by any of the authors. No approval of research ethics committees was required to accomplish this study because it was conducted on insects and crop plants.

### Consent to participate

The authors declare consent to participate in this article.

**Consent for publication**

The authors declare consent for the publication of this article.

**Table 1.** Formulations of synthetic fungicides and mycoinsecticides used in the compatibility studies.

| Active ingredients [concentration (g $L^{-1}$ or g $kg^{-1}$)] | Commercial brands | Formulation[1] | Tested dose[2] (ml or g $ha^{-1}$) | Manufacturer |
|---|---|---|---|---|
| ***Synthetic fungicides*** | | | | |
| Pyraclostrobin + epoxiconazole (260.0 + 160.0) | Abacus HC® | CS | 380.0 | Basf SE |
| Fluxapyroxad + pyraclostrobin (167.0 + 333.0) | Orkestra® | CS | 350.0 | Basf SE |
| Epoxiconazole + fluxapyroxad + pyraclostrobin (50.0 + 50.0 + 81.0) | Ativum® | EC | 1000.0 | Basf SE |
| Bixafen + prothioconazole + trifloxystrobin (125.0 + 175.0 + 150.0) | Fox® Xpro | CS | 500.0 | Bayer S/A |
| Azoxystrobin + difenoconazole (200.0 + 125.0) | Priori Top® | CS | 400.0 | Syngenta Proteção de Cultivos Ltda. |
| Trifloxystrobin + tebuconazole (100.0 + 200.0) | Nativo® | CS | 750.0 | Bayer S/A |
| Difenoconazole + pydiflumetofem (125.0 + 75.0) | Miravis Duo® | CS | 600.0 | Syngenta Proteção de Cultivos Ltda. |
| Trifloxystrobin + cyproconazole (375.0 + 160.0) | Sphere Max® | CS | 200.0 | Bayer S/A |
| Propiconazole + difenoconazole (250.0 + 250.0) | Score Flexi® | EC | 300.0 | Syngenta Proteção de Cultivos Ltda. |
| ***Mycoinsecticides*** | | | | |
| *Beauveria bassiana* (Simbi BB15 isolate) | FlyControl® | OD | 500.0 | Simbiose Ltda. |
| *Beauveria bassiana* (IBCB 66 isolate) | Bovenat® | WP | 500.0 | Bionat Soluções Biológicas Ltda. |
| *Cordyceps javanica* (Esalq-1296 isolate) | Octane® | CS | 500.0 | Koppert do Brasil Ltda. |

[1]Formulations: CS = concentrated suspension; EC = emulsifiable concentrate; OD = oil dispersion and WP = wettable powder.

**Table 2.** Colony diameter (mm), number of conidia (× $10^7$), and conidial germination (%) of isolate Esalq-1296 of *Cordyceps javanica* and isolates IBCB66 and Simbi BB15 of *Beauveria bassiana* exposed in a culture medium containing different synthetic fungicides after 12 days of incubation.

| Synthetic fungicides | Fungi isolates[1] | | |
|---|---|---|---|
| | Esalq-1296 | IBCB66 | Simbi BB15 |
| | Colony diameter (mm) | | |
| Pyraclostrobin + epoxiconazole | 0.00 ± 0.00 | 0.00 ± 0.00 | 0.00 ± 0.00 |
| Fluxapyroxad + pyraclostrobin | 0.00 ± 0.00 | 0.00 ± 0.00 | 0.00 ± 0.00 |
| Epoxiconazole + fluxapyroxad + pyraclostrobin | 0.00 ± 0.00 | 0.00 ± 0.00 | 0.00 ± 0.00 |
| Bixafem + prothioconazole + trifloxystrobin | 0.00 ± 0.00 | 0.00 ± 0.00 | 0.00 ± 0.00 |
| Azoxystrobin + difenoconazole | 0.00 ± 0.00 | 0.00 ± 0.00 | 0.00 ± 0.00 |
| Trifloxystrobin + tebuconazole | 0.00 ± 0.00 | 0.00 ± 0.00 | 0.00 ± 0.00 |
| Difenoconazole + pydiflumetofen | 0.00 ± 0.00 | 0.00 ± 0.00 | 0.00 ± 0.00 |
| Trifloxystrobin + ciproconazole | 0.00 ± 0.00 | 0.00 ± 0.00 | 0.00 ± 0.00 |
| Propiconazole +difenoconazole | 0.00 ± 0.00 | 0.00 ± 0.00 | 0.00 ± 0.00 |
| Negative control (distilled water) | 43.74 ± 1.10 | 46.73 ± 1.31 | 44.22 ± 1.80 |
| | Number of conidia (× $10^7$) | | |
| Pyraclostrobin + epoxiconazole | - | - | - |
| Fluxapyroxad + pyraclostrobin | - | - | - |
| Epoxiconazole + fluxapyroxad + pyraclostrobin | - | - | - |
| Bixafem + prothioconazole + trifloxystrobin | - | - | - |
| Azoxystrobin + difenoconazole | - | - | - |
| Trifloxystrobin + tebuconazole | - | - | - |
| Difenoconazole + pydiflumetofen | - | - | - |
| Trifloxystrobin + ciproconazole | - | - | - |
| Propiconazole +difenoconazole | - | - | - |
| Negative control (distilled water) | 535.31 ± 66.93 | 584.68 ± 52.60 | 265.50 ± 23.29 |
| | Conidial germination (%) | | |
| Pyraclostrobin + epoxiconazole | - | - | - |
| Fluxapyroxad + pyraclostrobin | - | - | - |
| Epoxiconazole + fluxapyroxad + pyraclostrobin | - | - | - |
| Bixafem + prothioconazole + trifloxystrobin | - | - | - |
| Azoxystrobin + difenoconazole | - | - | - |
| Trifloxystrobin + tebuconazole | - | - | - |
| Difenoconazole + pydiflumetofen | - | - | - |
| Trifloxystrobin + ciproconazole | - | - | - |
| Propiconazole +difenoconazole | - | - | - |
| Negative control (distilled water) | 98.63 ± 3.72 | 97.78 ± 2.97 | 98.62 ± 3.32 |

[1]Data not analyzed due to the absence of variance.

**Table 3**. *In vitro* toxicity classification of synthetic fungicides to isolate Esalq-1296 of *Cordyceps javanica* and isolates IBCB66 and Simbi BB15 of *Beauveria bassiana*.

| Synthetic fungicides | Commercial brands | Fungi isolates | | | | | |
|---|---|---|---|---|---|---|---|
| | | Esalq-1296 | | IBCB 66 | | Simbi BB15 | |
| | | *BI*[1] | Classification[2] | *BI*[1] | Classification[2] | *BI*[1] | Classification[2] |
| Pyraclostrobin + epoxiconazole | Abacus HC® | 0.00 | Incompatible | 0.00 | Incompatible | 0.00 | Incompatible |
| Fluxapyroxad + pyraclostrobin | Orkestra® | 0.00 | Incompatible | 0.00 | Incompatible | 0.00 | Incompatible |
| Epoxiconazole + fluxapyroxad + pyraclostrobin | Ativum® | 0.00 | Incompatible | 0.00 | Incompatible | 0.00 | Incompatible |
| Bixafem + prothioconazole + trifloxystrobin | Fox® Xpro | 0.00 | Incompatible | 0.00 | Incompatible | 0.00 | Incompatible |
| Azoxystrobin + difenoconazole | Priori Top® | 0.00 | Incompatible | 0.00 | Incompatible | 0.00 | Incompatible |
| Trifloxystrobin + tebuconazole | Nativo® | 0.00 | Incompatible | 0.00 | Incompatible | 0.00 | Incompatible |
| Difenoconazole + pydiflumetofen | Miravis Duo® | 0.00 | Incompatible | 0.00 | Incompatible | 0.00 | Incompatible |
| Trifloxystrobin + ciproconazole | Sphere Max® | 0.00 | Incompatible | 0.00 | Incompatible | 0.00 | Incompatible |
| Propiconazole +difenoconazole | Score Flexi® | 0.00 | Incompatible | 0.00 | Incompatible | 0.00 | Incompatible |

[1]*BI*: Biological index proposed by Rossi-Zalaf *et al*. (2008).

[2]Classification: incompatible ($0 \leq BI \leq 41$); moderately compatible ($42 \leq BI \leq 66$), and compatible ($BI > 66$).

**Table 4.** Colony-forming units (CFUs) of Esalq-1296 of *Cordyceps javanica* and isolates IBCB66 and Simbi BB15 of *Beauveria bassiana* exposed to different fungicides in tank mixtures in two exposure times (1.5 or 3 h).

| Synthetic fungicides | Esalq-1296 | | IBCB66 | | Simbi BB15 | |
|---|---|---|---|---|---|---|
| | 1.5 h | 3 h | 1.5 h | 3 h | 1.5 h | 3 h |
| Pyraclostrobin + epoxiconazole | 0.00 ± 0.00 | 0.00 ± 0.00 | 0.00 ± 0.00 | 0.00 ± 0.00 | 0.00 ± 0.00 | 0.00 ± 0.00 |
| Fluxapyroxad + pyraclostrobin | 0.00 ± 0.00 | 0.00 ± 0.00 | 0.00 ± 0.00 | 0.00 ± 0.00 | 0.00 ± 0.00 | 0.00 ± 0.00 |
| Epoxiconazole + fluxapyroxad + pyraclostrobin | 0.00 ± 0.00 | 0.00 ± 0.00 | 0.00 ± 0.00 | 0.00 ± 0.00 | 0.00 ± 0.00 | 0.00 ± 0.00 |
| Bixafem + prothioconazole + trifloxystrobin | 0.00 ± 0.00 | 0.00 ± 0.00 | 0.00 ± 0.00 | 0.00 ± 0.00 | 0.00 ± 0.00 | 0.00 ± 0.00 |
| Azoxystrobin + difenoconazole | 0.00 ± 0.00 | 0.00 ± 0.00 | 0.00 ± 0.00 | 0.00 ± 0.00 | 0.00 ± 0.00 | 0.00 ± 0.00 |
| Trifloxystrobin + tebuconazole | 0.00 ± 0.00 | 0.00 ± 0.00 | 0.00 ± 0.00 | 0.00 ± 0.00 | 0.00 ± 0.00 | 0.00 ± 0.00 |
| Difenoconazole + pydiflumetofen | 0.00 ± 0.00 | 0.00 ± 0.00 | 0.00 ± 0.00 | 0.00 ± 0.00 | 0.00 ± 0.00 | 0.00 ± 0.00 |
| Trifloxystrobin + ciproconazole | 0.00 ± 0.00 | 0.00 ± 0.00 | 0.00 ± 0.00 | 0.00 ± 0.00 | 0.00 ± 0.00 | 0.00 ± 0.00 |
| Propiconazole +difenoconazole | 0.00 ± 0.00 | 0.00 ± 0.00 | 0.00 ± 0.00 | 0.00 ± 0.00 | 0.00 ± 0.00 | 0.00 ± 0.00 |
| Negative control (distilled water) | 234.00 ± 3.31 | 239.40 ± 4.41 | 339.70 ± 3.99 | 337.00 ± 4.12 | 189.00 ± 4.79 | 192.00 ± 5.19 |

**Table 5.** Colony-forming units (CFUs) of commercial mycoinsecticides Octane® (isolate Esalq-1296 from *Cordyceps javanica*) and FlyControl® (isolate Simbi BB15 from *Beauveria bassiana*) exposed to different fungicides in tank mixtures at two exposure times (1.5 and 3 h).

| Synthetic fungicides | Exposure times (h)[1] | |
|---|---|---|
| | 1.5 | 3 |
| FlyControl® (*Beauveria bassiana* isolate Simbi BB15) | | |
| Pyraclostrobin + epoxiconazole | 0.00 ± 0.00* | 0.00 ± 0.00* |
| Fluxapyroxad + pyraclostrobin | 6.20 ± 0.52 cdA | 0.00 ± 0.00 *B |
| Epoxiconazole + fluxapyroxad + pyraclostrobin | 2.40 ± 1.04 dA | 0.00 ± 0.00 *B |
| Bixafem + prothioconazole + trifloxystrobin | 186.60 ± 3.38 bA | 140.80 ± 4.10 dB |
| Azoxystrobin + difenoconazole | 21.00 ± 1.17 cA | 0.00 ± 0.00 *B |
| Trifloxystrobin + tebuconazole | 120.80 ± 4.47 bA | 72.20 ± 2.29 eB |
| Difenoconazole + pydiflumetofen | 321.80 ± 8.17 abA | 180.40 ± 5.52 cB |
| Trifloxystrobin + ciproconazole | 26.20 ± 1.48 cA | 0.00 ± 0.00 *B |
| Propiconazole +difenoconazole | 477.40 ± 8.51 abA | 450.20 ± 5.50 bB |
| Negative control (distilled water) | 678.40 ± 12.83 aA | 676.60 ± 9.56 aA |
| Fungicides (LRT= 4115.90; Df = 4, 49; $p < 0.01$) | | |
| Interaction between fungicides × exposure times (LRT = 533.41; Df = 8, 89; $p < 0.01$) | | |
| Octane® (*Cordyceps javanica* isolate Esalq-1296) | | |
| Pyraclostrobin + epoxiconazole | 0.00 ± 0.00* | 0.00 ± 0.00* |
| Fluxapyroxad + pyraclostrobin | 0.00 ± 0.00* | 0.00 ± 0.00* |
| Epoxiconazole + fluxapyroxad + pyraclostrobin | 0.00 ± 0.00* | 0.00 ± 0.00* |
| Bixafem + prothioconazole + trifloxystrobin | 1.00 ± 0.89 dA | 0.00 ± 0.00 *B |
| Azoxystrobin + difenoconazole | 0.00 ± 0.00* | 0.00 ± 0.00* |
| Trifloxystrobin + tebuconazole | 3.60 ± 1.80 dA | 0.00 ± 0.00 *B |
| Difenoconazole + pydiflumetofen | 542.40 ± 7.65 bA | 325.20 ± 7.47 bB |
| Trifloxystrobin + ciproconazole | 0.00 ± 0.00* | 0.00 ± 0.00* |
| Propiconazole +difenoconazole | 179.40 ± 6.23 cA | 96.80 ± 5.04 cB |
| Negative control (distilled water) | 736.00 ± 9.01 aA | 733.80 ± 10.47 aA |
| Fungicides (LRT= 2839.60; Df = 2, 29; $p < 0.01$) | | |
| Interaction between fungicides × exposure times (LRT = 252.18; Df = 4, 49; $p < 0.01$) | | |

[1]Data (mean ± SE) followed by the same letter in the columns do not differ significantly (GLM with Poisson distribution, followed by *post hoc* likelihood ratio test; $p > 0.05$).

*Treatment excluded from the analysis because there was no variability.

**Table 6.** Physical-chemical compatibility of mixtures of different synthetic fungicides registered for management of maize diseases and mycoinsecticides registered for *Dalbulus maidis* management in Brazil and tested according to criteria of dynamic assays proposed by the Brazilian Association of Technical Standards (ABNT NBR 13875:2014).

| Treatments | pH | CE (µS/cm) | ho | fl | sd | ps | pl | os | cf | cr | fo | pl* |
|---|---|---|---|---|---|---|---|---|---|---|---|---|
| **Synthetic fungicides** | | | | | | | | | | | | |
| Pyraclostrobin + epoxiconazole | 5.92 | 22.58 | P | A | A | A | A | A | A | A | A | A |
| Fluxapyroxad + pyraclostrobin | 5.72 | 35.10 | P | A | A | A | A | A | A | A | A | A |
| Epoxiconazole + fluxapyroxad + pyraclostrobin | 4.74 | 35.80 | P | A | A | A | A | A | A | A | A | A |
| Bixafem + prothioconazole + trifloxystrobin | 5.47 | 18.97 | P | A | A | A | A | A | A | A | A | A |
| Azoxystrobin + difenoconazole | 5.42 | 16.39 | P | A | A | A | A | A | A | A | A | A |
| Trifloxystrobin + tebuconazole | 7.07 | 38.40 | P | A | A | A | A | A | A | A | A | A |
| Difenoconazole + pydiflumetofen | 5.35 | 9.27 | P | A | A | A | A | A | A | A | A | A |
| Trifloxystrobin + ciproconazole | 5.64 | 6.46 | P | A | A | A | A | A | A | A | A | A |
| Propiconazole +difenoconazole | 6.27 | 38.70 | P | A | A | A | A | A | A | A | A | A |
| **Mycoinsecticides** | | | | | | | | | | | | |
| Fly Control® | 5.39 | 6.58 | P | A | A | A | A | A | A | A | A | A |
| Bovenat® | 5.48 | 9.09 | P | A | A | A | A | A | A | A | A | A |
| Octane® | 5.50 | 8.09 | P | A | A | A | A | A | A | A | A | A |
| Água | 5.40 | 3.95 | - | - | - | - | - | - | - | - | - | - |
| **Synthetic fungicides + mycoinsecticides** | | | | | | | | | | | | |
| Pyraclostrobin + epoxiconazole + Fly Control® | 5.93 | 22.84 | P | A | A | A | A | A | A | A | A | A |
| Fluxapyroxad + pyraclostrobin + Fly Control® | 6.01 | 36.60 | P | A | A | A | A | A | A | A | A | A |
| Epoxiconazole + fluxapyroxad + pyraclostrobin + Fly Control® | 4.83 | 37.50 | A | A | A | P | A | A | A | A | A | A |
| Bixafem + prothioconazole + trifloxystrobin + Fly Control® | 5.86 | 20.77 | A | A | A | P | A | A | A | A | A | A |
| Azoxystrobin + difenoconazole + Fly Control® | 5.70 | 18.19 | P | A | A | A | A | A | A | A | A | A |
| Trifloxystrobin + tebuconazole + Fly Control® | 6.90 | 41.50 | A | A | A | P | A | A | A | A | A | A |
| Difenoconazole + pydiflumetofen + Fly Control® | 5.45 | 11.16 | P | A | A | A | A | A | A | A | A | A |
| Trifloxystrobin + ciproconazole + Fly Control® | 5.92 | 8.80 | P | A | A | A | A | A | A | A | A | A |
| Propiconazole +difenoconazole + Fly Control® | 6.34 | 40.30 | A | A | A | P | A | A | A | A | A | A |
| Pyraclostrobin + epoxiconazole + Octane® | 6.19 | 34.40 | P | A | A | A | A | A | A | A | A | A |
| Fluxapyroxad + pyraclostrobin + Octane® | 6.09 | 49.70 | P | A | A | A | A | A | A | A | A | A |
| Epoxiconazole + fluxapyroxad + pyraclostrobin + Octane® | 5.49 | 42.90 | A | A | A | P | A | A | A | A | A | A |
| Bixafem + prothioconazole + trifloxystrobin + Octane® | 6.04 | 31.10 | P | A | A | A | A | A | A | A | A | A |
| Azoxystrobin + difenoconazole + Octane® | 5.85 | 32.20 | P | A | A | A | A | A | A | A | A | A |
| Trifloxystrobin + tebuconazole + Octane® | 6.76 | 51.00 | P | A | A | A | A | A | A | A | A | A |
| Difenoconazole + pydiflumetofen + Octane® | 5.94 | 21.49 | P | A | A | A | A | A | A | A | A | A |
| Trifloxystrobin + ciproconazole + Octane® | 6.11 | 19.35 | P | A | A | A | A | A | A | A | A | A |
| Propiconazole +difenoconazole + Octane® | 6.47 | 62.20 | P | A | A | A | A | A | A | A | A | A |
| Pyraclostrobin + epoxiconazole + Bovenat® | 5.83 | 27.66 | P | A | A | A | A | A | A | A | A | A |
| Fluxapyroxad + pyraclostrobin + Bovenat® | 5.87 | 41.10 | P | A | A | A | A | A | A | A | A | A |
| Epoxiconazole + fluxapyroxad + pyraclostrobin + Bovenat® | 4.68 | 76.30 | P | A | A | A | A | A | A | A | A | A |
| Bixafem + prothioconazole + trifloxystrobin + Bovenat® | 5.68 | 22.84 | P | A | A | A | A | A | A | A | A | A |
| Azoxystrobin + difenoconazole + Bovenat® | 5.76 | 25.52 | P | A | A | A | A | A | A | A | A | A |
| Trifloxystrobin + tebuconazole + Bovenat® | 6.87 | 53.60 | P | A | A | A | A | A | A | A | A | A |
| Difenoconazole + pydiflumetofen + Bovenat® | 5.84 | 24.37 | P | A | A | A | A | A | A | A | A | A |
| Trifloxystrobin + ciproconazole + Bovenat® | 5.99 | 20.80 | P | A | A | A | A | A | A | A | A | A |
| Propiconazole +difenoconazole + Bovenat® | 6.28 | 51.40 | P | A | A | A | A | A | A | A | A | A |

P: presence or A: absence (from assessment criteria). pH after agitation; EC: electrical conductivity; ho: homogeneity; fl: flocculation; sd: sedimentation; ps: phase separation; pg: presence of groats; os: oil separation; cf: crystal formation; cm: cream; fo: foam; pg*: presence of groats (sieve de 149 µm).

**Table 7.** Percentages of *Dalbulus maidis* mortality and proportion of extrudates adults as well as the number of conidia produced in each insect cadaver exposed to mixtures containing fungicides, mycoinsecticides, and the mixture of both.

| Treatment | % Mortality[1] | % Extrusion[2] | Number of conidia ($\times 10^7$)[3] |
|---|---|---|---|
| Pyraclostrobin + epoxiconazole | 6.00 ± 2.45 b | 0.00 | - |
| Fluxapyroxad + pyraclostrobin | 2.00 ± 2.00 b | 0.00 | - |
| Epoxiconazole + fluxapyroxad + pyraclostrobin | 26.00 ± 5.10 ab | 0.00 | - |
| Bixafem + prothioconazole + trifloxystrobin | 2.00 ± 2.14 b | 0.00 | - |
| Azoxystrobin + difenoconazole | 2.00 ± 2.00 b | 0.00 | - |
| Trifloxystrobin + tebuconazole | 6.00 ± 2.45 b | 0.00 | - |
| Difenoconazole + pydiflumetofen | 4.00 ± 2.46 b | 0.00 | - |
| Trifloxystrobin + ciproconazole | 8.00 ± 3.74 b | 0.00 | - |
| Propiconazole +difenoconazole | 4.00 ± 2.45 b | 0.00 | - |
| Pyraclostrobin + epoxiconazole + Octane® | 28.00 ± 3.34 ab | 0.00 | - |
| Fluxapyroxad + pyraclostrobin + Octane® | 36.00 ± 6.06 ab | 0.00 | - |
| Epoxiconazole + fluxapyroxad + pyraclostrobin + Octane® | 54.00 ± 5.37 a | 0.00 | - |
| Bixafem + prothioconazole + trifloxystrobin + Octane® | 20.00 ± 2.82 ab | 0.00 | - |
| Azoxystrobin + difenoconazole + Octane® | 8.00 ± 3.34 b | 0.00 | - |
| Trifloxystrobin + tebuconazole + Octane® | 24.00 ± 5.36 ab | 0.00 | - |
| Difenoconazole + pydiflumetofen + Octane® | 22.00 ± 4.38 ab | 0.00 | - |
| Trifloxystrobin + ciproconazole + Octane® | 14.00 ± 4.56 b | 0.00 | - |
| Propiconazole +difenoconazole + Octane® | 10.00 ± 2.82 b | 0.00 | - |
| Pyraclostrobin + epoxiconazole + FlyControl® | 48.00 ± 6.57 a | 0.00 | - |
| Fluxapyroxad + pyraclostrobin + FlyControl® | 52.00 ± 4.38 a | 0.00 | - |
| Epoxiconazole + fluxapyroxad + pyraclostrobin + FlyControl® | 64.00 ± 6.06 a | 0.00 | - |
| Bixafem + prothioconazole + trifloxystrobin + FlyControl® | 22.00 ± 5.21 ab | 0.00 | - |
| Azoxystrobin + difenoconazole + FlyControl® | 26.00 ± 6.07 ab | 0.00 | - |
| Trifloxystrobin + tebuconazole + FlyControl® | 34.00 ± 11.52 ab | 0.00 | - |
| Difenoconazole + pydiflumetofen + FlyControl® | 26.00 ± 9.20 ab | 0.00 | - |
| Trifloxystrobin + ciproconazole + FlyControl® | 42.00 ± 4.38 ab | 0.00 | - |
| Propiconazole +difenoconazole + FlyControl® | 24.00 ± 2.19 ab | 0.00 | - |
| *C. javanica* Esalq-1296 (Octane®) | 22.00 ± 1.78 ab | 27.27 | 1.49 ± 0.09 a |
| *B. bassiana* Simbi BB 15 (FlyControl®) | 28.00 ± 3.35 ab | 21.43 | 0.69 ± 0.04 b |
| Bifenthrin + carbosulfan (positive control) | 100.00 ± 0.00* | 0.00 | - |
| Control (negative control) | 2.00 ± 1.78 b | 0.00 | - |
| LRT | 262.21 | | 16.89 |
| Df | 29, 120 | | 1, 4 |
| *P* | < 0.01 | | < 0.01 |

[1]Data (mean ± SE) followed by the same letter in a column do not differ significantly (GLM with beta binomial distribution, followed by *post hoc* likelihood ratio test; $p > 0.05$).
[2]Percentage of extruded insects in relation to the total number of deaths.
[3]Data (mean ± SE) followed by the same letter in a column do not differ significantly (GLM with negative binomial distribution, followed by *post hoc* likelihood ratio test; $p > 0.05$).
[*]Treatment excluded from the analysis because there was no variability.

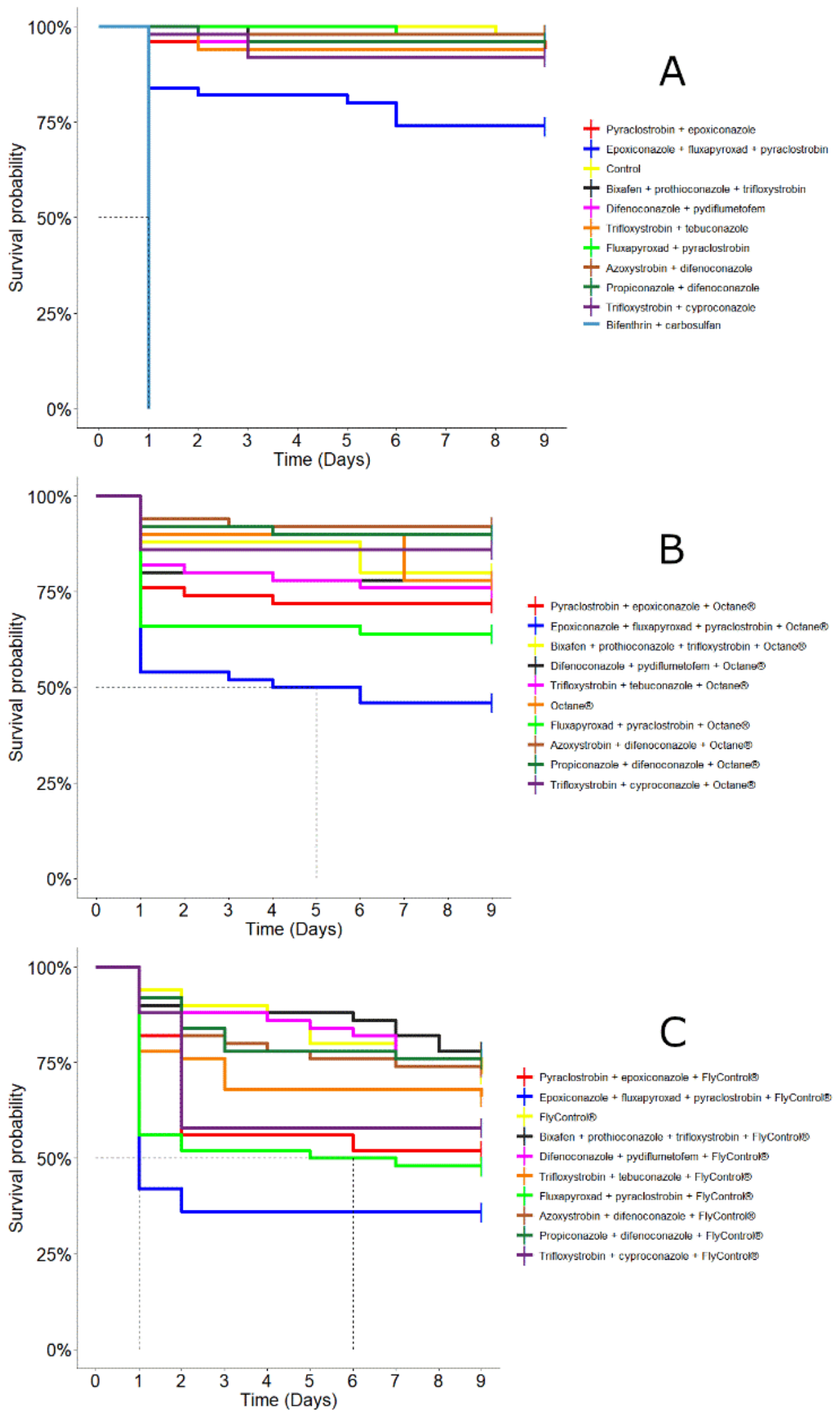

**Figure 1.** Survival probability of *Dalbulus maidis* adults exposed to fungicides applied alone (A) or in binary mixtures with the commercial mycoinsecticides Octane® (B) and FlyControl® (C). *Dotted lines correspond to the median lethal time of the respective treatment.

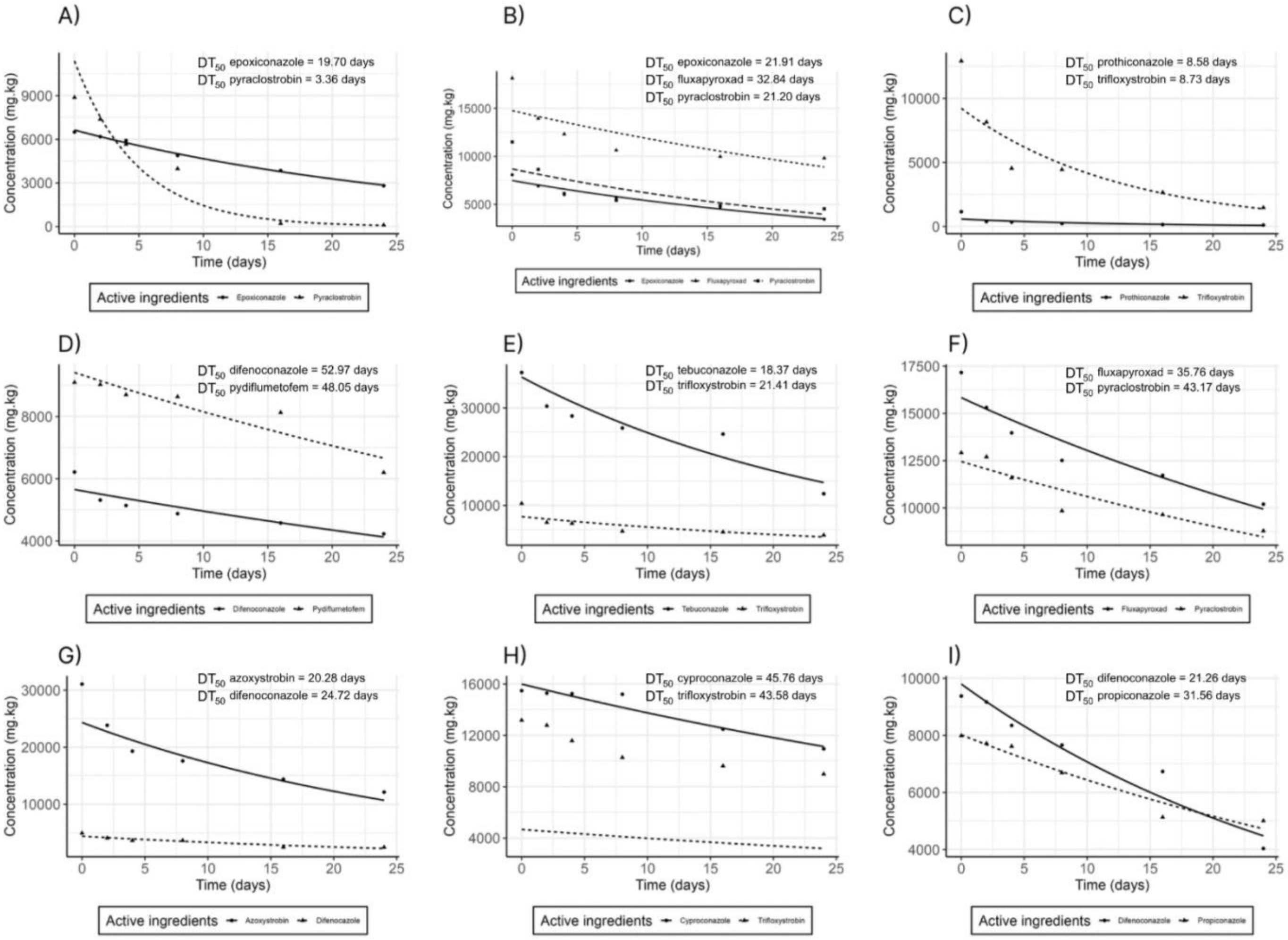

**Figure 2.** Pesticides residues in maize leaves following application. The figures represent commercial synthetic fungicides: (A) Abacus HC®; (B) Ativum®; (C) Fox Xpro®; (D) Miravis Duo®; (E) Nativo®; (F) Orkestra®; (G) Priori Top®; (H) Sphere Max® and (I) Score Flexi®. Data points represent the residue concentration at different sampling times. Lines represent the degradation kinetics of the corresponding treatment, fitted by an exponential regression in a simple first-order (SFO) model.